\documentstyle[prl,aps,multicol]{revtex}
\begin{document}
\draft
\preprint{HD-THEP-00-37}
\title{Unique Translation between Hamiltonian Operators and Functional Integrals}
\author{Tim Gollisch\cite{mailgoll} and Christof Wetterich\cite{mailwett}}
\address{
Institut f\"ur Theoretische Physik,
Universit\"at Heidelberg,
Philosophenweg 16, 69120 Heidelberg,
Germany
}
\date{\today}
\maketitle
\begin{abstract}
A careful treatment of the discretization errors in the path integral formulation of quantum mechanics leads to a unique prescription for the translation from the Hamiltonian to the action in the functional integral.  An example is given by an interaction quadratic in the occupation number, characteristic for manybody bosonic systems.  As a result, the term linear in the occupation number (chemical potential) receives a correction as compared to the usual formulation based on coherent states.  A perturbative calculation supports the relevance of this correction.
\end{abstract}

\begin{multicols}{2}

\section{Introduction}

Many modern techniques in particle and statistical physics rely on theories given in the form of functional integrals, while the basic description of the system is often easier and more intuitively formulated on the operator level.  The transition between these two languages has been given much attention (see e.g.\ \cite{Feynman,Garrod,Casher,RevzenSchulman,Fanelli,ChristLee,Peeters,deBoer}) and is recognized as a mathematically complicated issue if the Hamiltonian cannot be separated into a momentum and a location dependent part.  Nevertheless, calculations based on functional integral formulations have been very successful in applications, for instance in the context of the renormalization group in statistical physics.

In order to interpret the parameters which appear in the functional integral correctly, it is necessary to connect them to the parameters of the Hamiltonian operator ${\cal H}$.  This translation between the parameters in the functional integral and the Hamiltonian has to be unique since physical quantities can be computed both ways.  (If necessary, manybody models have to be regularized such that the expectation values of ${\cal H}$ and other operators are well defined.  Typically, this can be done by discretization on a lattice.)

For an ensemble of interacting bosons of mass $m$ with the Hamiltonian
\begin{equation} \label{eq:H_B}
{\cal H}\! = \!\!\sum_q \!\!\left(\!\frac{q^2}{2m}\!-\!\mu\!\right)\!a_q^\dag a_q
\!+\!\frac{1}{2}\!\!\!\sum_{q_1,q_2,q_3}\!\!\!\!\!a^\dag_{q_1\!-\!q_3}a^\dag_{q_2\!+\!q_3}v(q_3)a_{q_1}a_{q_2},
\end{equation}
the partition function is written as a functional integral over a complex field $\varphi_q(\tau)$
\begin{equation}
Z=\int\!{\cal D} \varphi \;\exp\! \left\{ -{\cal S}\left[\varphi,\varphi^*\right] \right\}
\end{equation}
with the action
\begin{eqnarray} \nonumber
{\cal S}\!\left[\varphi,\varphi^*\right]&=& \int \!d\tau \Big[\sum_q \varphi_q^*\left(\frac{q^2}{2m}-\tilde{\mu}+\partial_\tau\right)\varphi_q\\
& &+\frac{1}{2}\!\!\sum_{q_1,q_2,q_3}\!\!\varphi_{q_1-q_3}^*\varphi_{q_2+q_3}^*\tilde{v}(q_3)\varphi_{q_1}\varphi_{q_2}\Big]. \label{eq:cl_ac_bosonic}
\end{eqnarray}
Usually, $\tilde{\mu}$ is interpreted as the chemical potential $\mu$ and $\tilde{v}(q)$ as the interaction potential $v(q)$ \cite{Casher,Stoof,BaymZJ}.

The intuitive derivation makes use of the coherent states introduced by Glauber \cite{Glauber} and can be found in Ref.~\cite{Casher}.  It has been suggested earlier, though, that the discretization errors, namely the $(\Delta \tau)^2$--corrections in the exponent, are of importance and cannot be discarded thoughtlessly \cite{Schulman}.  In this work, we present an approach which controls these higher order corrections.

As a consequence, the parameter $\tilde{\mu}$ obtains a correction from the mathematical manipulations and is to be interpreted as $\mu+\Omega$, where $\Omega$ is a constant depending on $v(q)$.
The derivation we present here leads to a unique relationship between the Hamiltonian and the functional integral.

This issue is related to the work of Christ and Lee \cite{ChristLee} who obtain extra terms in the quantum Yang--Mills Hamiltonian from a gauge transformation on the operator level.  They claim that this is connected to non--negligible terms of higher order in the discretization parameter $\epsilon$ in the functional integral.  The existence of such possible subtleties in the discretization is our starting point.  We propose a formulation that avoids those unpleasant higher order contributions.

There has been recent significant improvement in the understanding of phase space integrals by de Boer, Peeters, Skenderis, and van Nieuwenhuizen \cite{Peeters,deBoer}, see also \cite{Waldron}.  They derive the action and Feynman rules for Hamiltonian operators in curved spacetime which are at most quadratic in $\hat{p}$.  This is applied to the non--linear sigma model.  In this work we are interested in the statistical physics applications of functional integrals where the interactions typically involve higher powers of $\hat{p}$.  We therefore take a different approach, which allows us to treat momentum and location operators on equal footing.  (There is no restriction to Hamiltonians at most quadratic in $\hat{p}$).  Our simple treatment leads (for flat spacetime) to a perhaps unexpected result, namely a correction in the action which we can connect to experimental findings such as the critical chemical potential at the lambda--transition of helium--4.

\section{Functional integral in quantum mechanics}

The mathematical issues of path integration are most easily studied in one dimensional quantum mechanics, which is a finite theory.  The generalization of the results from this toy model to quantum field theory is straightforward and given below (\ref{helium}).
We consider the canonical partition function with the Hamiltonian
\begin{equation}
{\cal H}=ma^\dag a + \frac{\lambda}{2}a^\dag a^\dag a a.
\label{eq:H_anh_osc}
\end{equation}
We introduce location and momentum operators
$\hat{x}=\left(a^\dag+a\right)/\sqrt{2}\hspace{2mm}$ and 
$\hat{p}=i\left(a^\dag-a\right)/\sqrt{2},$
which obey the usual commutation relation $\left[\hat{x},\hat{p}\right]=i$.
Bringing the interaction term into a symmetric ordering yields
\begin{mathletters} \label{eq:H_xp}
\begin{equation}
{\cal H}={\cal H_A}+{\cal H_B}-\frac{m}{2}+\frac{3\lambda}{8}
\end{equation}
with
\begin{eqnarray}
{\cal H_A}&=&\frac{1}{4}\left(m\!-\!\lambda\right)\left(\hat{p}^2\!+\!\hat{x}^2\right)
+\frac{\lambda}{16}\left(\hat{x}^4\!+\!2\hat{p}^2\hat{x}^2\!+\!\hat{p}^4\right), \\
{\cal H_B}&=&\frac{1}{4}\left(m\!-\!\lambda\right)\left(\hat{p}^2\!+\!\hat{x}^2\right)
+\frac{\lambda}{16}\left(\hat{x}^4\!+\!2\hat{x}^2\hat{p}^2\!+\!\hat{p}^4\right).
\end{eqnarray}
\end{mathletters}
The term $-\frac{1}{2}\lambda\left(\hat{x}^2+\hat{p}^2\right)$ as well as the constant term $-\frac{m}{2}+\frac{3\lambda}{8}$ arise from the use of the commutators in the manipulation of (\ref{eq:H_anh_osc}).  The trivial constant term will be neglected for notational simplicity.

This particular ordering is a suitable starting point for the formulation of the functional integral.  We will show that it avoids unpleasant discretization corrections which would be present for other formulations like the one based on coherent states.

We denote the location and momentum eigenstates by $|x\rangle$ and $|p\rangle$ and follow the standard procedure for writing the partition function as a functional integral (with $\epsilon\equiv\beta /(2N)$ and $x_{2N}\equiv x_0$):
\begin{eqnarray}\nonumber
Z
&=& \mbox{Tr}\; e^{-\beta {\cal H}} = Z_N \\ \nonumber
&=&\!\!\!\int\!\!\!\bigg(\prod_{a=0}^{N-1}dx_{2a}\!\!\bigg)\!\!
\langle x_{2N} | e^{-2\epsilon{\cal H}} |x_{2N\!-\!2}\rangle \ldots \langle x_2| e^{-2\epsilon{\cal H}}|x_0\rangle. \\
\end{eqnarray}
In expanding the exponentials for small $\epsilon$, we will take special care of the $\epsilon^2$--contributions.
The matrix elements are evaluated by inserting further location and momentum eigenstates
\begin{eqnarray} \nonumber
\lefteqn{\langle x_{k+2}|1-2\epsilon{\cal H}+2\epsilon^2{\cal H}^2|x_k\rangle }\\ \nonumber
&&= \int d p_{k+1}\; d x_{k+1}\; d p_k \\ \nonumber
&&\times\big[\langle x_{k+2}|p_{k+1}\rangle\langle p_{k+1}|x_{k+1}\rangle\langle x_{k+1}|p_k\rangle\langle p_k|x_k\rangle \\ \nonumber
&&-\epsilon\,\big(\!\langle x_{k+2}|p_{k+1}\rangle\langle p_{k+1}|{\cal H_A}|x_{k+1}\rangle\langle x_{k+1}|p_k\rangle\langle p_k|x_k\rangle \\ \nonumber
&&\;\;\;\;+\langle x_{k+2}|p_{k+1}\rangle\langle p_{k+1}|x_{k+1}\rangle\langle x_{k+1}|p_k\rangle\langle p_k|{\cal H_A}|x_k\rangle \\ \nonumber
&&\;\;\;\;+\langle x_{k+2}|{\cal H_B}|p_{k+1}\rangle\langle p_{k+1}|x_{k+1}\rangle\langle x_{k+1}|p_k\rangle\langle p_k|x_k\rangle \\ \nonumber
&&\;\;\;\;+\langle x_{k+2}|p_{k+1}\rangle\langle p_{k+1}|x_{k+1}\rangle\langle x_{k+1}|{\cal H_B}|p_k\rangle\langle p_k|x_k\rangle\big) \\ \nonumber
&&+2\epsilon^2\big(\!\langle x_{k+2}|p_{k+1}\rangle\langle p_{k+1}|{\cal H_A}|x_{k+1}\rangle\langle x_{k+1}|p_k\rangle\langle p_k|{\cal H_A}|x_k\rangle \\ \nonumber
&&\;\;\;\;+\!\langle x_{k+2}|{\cal H_B}|p_{k+1}\rangle\langle p_{k+1}|x_{k+1}\rangle\langle x_{k+1}|{\cal H_B}|p_k\rangle\langle p_k|x_k\rangle \\ \nonumber
&&\;\;\;\;+\!\langle x_{k+2}|p_{k+1}\rangle\langle p_{k+1}|{\cal H_A}|x_{k+1}\rangle\langle x_{k+1}|{\cal H_B}|p_k\rangle\langle p_k|x_k\rangle \\ \nonumber
&&\;\;\;\;+\!\langle x_{k+2}|{\cal H_B}|p_{k+1}\rangle\langle p_{k+1}|{\cal H_A}|x_{k+1}\rangle\langle x_{k+1}|p_k\rangle\langle p_k|x_k\rangle\!\big)\big]. \\
\end{eqnarray}
In the above expression, all matrix elements are trivial to evaluate, and we see that the $\epsilon^2$--term is almost exactly the one we need for reconverting the expansion into an exponential.  All we need to do is adjust some of the indices ($k+2\leftrightarrow k+1 \leftrightarrow k$) in this term, but this gives a correction of ${\cal O}(\epsilon^3)$ (as will be evident in (\ref{eq:dft_xp})).  Hence we can write
\begin{eqnarray} \nonumber
Z_N&=&\int\frac{dx_0}{\sqrt{2\pi}}\ldots\frac{dx_{2\!N\!-\!1}}{\sqrt{2\pi}}\frac{dp_0}{\sqrt{2\pi}}\ldots\frac{dp_{2\!N\!-\!1}}{\sqrt{2\pi}} \\ \nonumber
& & \times \exp\Bigg\{ \! \sum_{k=0}^{2\!N\!-\!1} \! \Big[ ip_k\left(x_{k+1}\!-\!x_k\right)-\frac{\epsilon}{2}\left(m\!-\!\lambda\right)\left(x_k^2\!+\!p_k^2\right) \\
& &-\frac{\epsilon\lambda}{8}\left(x_k^4+p_k^2x_k^2+x_{k+1}^2p_k^2+p_k^4\right)\Big]+{\cal O}(\epsilon^3)\Bigg\}.
\end{eqnarray}
The crucial question is whether $Z_N$ has a well defined limit for $N\rightarrow\infty$ such that the corrections ${\cal O}(\epsilon^3)$ can be neglected.  Since this is closely related to the issue of rapidly varying $x$ and $p$ (as functions of $\tau=\beta k/(2N)$), it is useful to perform a Fourier transform.
The definition of the Matsubara--modes $\tilde{x}_n$, $\tilde{p}_n$
\begin{mathletters} \label{eq:dft_xp}
\begin{eqnarray}
x_k&=&\sum_{n=-N}^{N} e^{2\pi ink/(2N)}\tilde{x}_n, \hspace{1.5mm} \\
p_k&=&\sum_{n=-N}^{N} e^{2\pi in(k+1/2)/(2N)}\tilde{p}_n
\end{eqnarray}
\end{mathletters}
(with $\tilde{x}_n^*=\tilde{x}_{-n}$, $\tilde{p}_n^*=\tilde{p}_{-n}$ and $\tilde{x}_0$, $\tilde{p}_0$, $\tilde{x}_N$, $\tilde{p}_N$ real)
corresponds in the imaginary time language to taking the variables $p(\tau)$ at locations between the $x(\tau)$.  From (\ref{eq:dft_xp}), we see that shifting the $k$--index introduces a factor of $1+{\cal O}(1/N)$.
We also turn to the more convenient language of complex fields by substituting
$\tilde{x}_n=\left(\varphi^*_{-n}+\varphi_n\right)/\sqrt{2}$ and
$\tilde{p}_n=i\left(\varphi^*_{-n}-\varphi_n\right)/\sqrt{2}.$
This yields
\begin{equation} \label{eq:func_int_def}
Z_N=\int {\cal D}\varphi\;\exp\left[-({\cal S}+\Delta{\cal S})+{\cal O}(1/N^2)\right] 
\end{equation}
with
\begin{eqnarray} \nonumber
{\cal S}&=&
\sum_n \left[ 2\pi i n \varphi^*_n \varphi_n +\beta\left(m-\lambda\right) \varphi^*_n \varphi_n \right] \\
&&+\frac{\beta\lambda}{2}\!\!\sum_{n_1,n_2,n}\!\! \varphi^*_{n_1-n} \varphi^*_{n_2+n} \varphi_{n_1} \varphi_{n_2}, \\ \nonumber
\Delta{\cal S}&=&\sum_n\left[4iN \sin\left(\frac{\pi n}{2N}\right)-2\pi i n\right]\varphi_n^*\varphi_n \\ \nonumber
&&+\frac{\beta\lambda}{16}\!\!\!\sum_{n_1,n_2,n}\!\!\left[1\!-\!\cos\left(\frac{\pi(n_1+n_2)}{2N}\right)\!\right]\!(\varphi^*_{\!-\!n_1\!+\!n}\!-\!\varphi_{n_1\!-\!n}) \\
&&\times (\varphi^*_{\!-\!n_2\!-\!n}-\varphi_{n_2\!+\!n})(\varphi^*_{\!-\!n_2}-\varphi_{n_2})(\varphi^*_{\!-\!n_1}-\varphi_{n_1}).
\end{eqnarray}
We notice that the coupling constants in the $\Delta{\cal S}$--correction are $\propto{\cal O}(1/N^2)$.  (The $n$--dependence is not important in this respect.  Up to a constant prefactor of $Z$, the contributions of the high $n$ Matsubara--modes are effectively cut off by suppression factors $\propto n^{-2}$ from the squared propagator, which arise from the first term in ${\cal S}$.)  Eq.~(\ref{eq:func_int_def}) is therefore particularly suited for an unequivocal definition of the functional integral as the limit $N\rightarrow\infty$ is taken.  We extend the $n$--summation from $-\infty$ to $\infty$ in the action ${\cal S}$ and drop $\Delta {\cal S}$ as well as the other ${\cal O}(1/N^2)$--corrections.

In the ``imaginary time language'' (using $\varphi(\tau)=\sum_n e^{2 \pi i n \tau / \beta} \varphi_n$), the action corresponds to
\begin{equation}
{\cal S}\!=\!\! \int_{-\beta/2}^{\beta/2}\! d\tau \Big[ \varphi^*(\tau)\left(\partial_\tau\!+\!m\!-\!\lambda\right)\varphi(\tau)
+\frac{\lambda}{2}|\varphi(\tau)|^4\!\Big].
\end{equation}
In this simple case, the only difference to the naive use of coherent states
is the shift of $\lambda$ in the mass term (besides constant terms).

Only the symmetric ordering of (\ref{eq:H_xp}) avoids unpleasant $1/N$--corrections.  This criterion leads to a unique translation into a continuous functional integral.  We can also take the perspective from the problem of operator ordering \cite{Kerner_Sutcliffe,Cohen,Mizrahi,Dowker}.  To think of the easiest example, the different operators $\hat{x}\hat{p}$ and $\hat{p}\hat{x}$ give the same zeroth order contribution in the action.  This leads to an ambiguous functional integral if higher order corrections are not taken into account.  The difference between these operators must occur elsewhere in the functional integral, and we see that it precisely appears in the $1/N$--terms.  These can give finite contributions if the evaluation of the functional integral preceeds the limit $N\rightarrow\infty$.  In the transition to the continuous functional integral, they would thus be erroneously neglected.


\section{A simple test}

In order to check our assertion that the quadratic term should be $m-\lambda$ instead of $m$, we calculate the thermal equilibrium value of the momentum squared from the functional integral in first order perturbation theory.  This is compared to a direct evaluation from ordinary quantum mechanics.  The two conclusions we draw from this will be that the functional integral gives finite results and therefore needs no regularization and that only the quadratic term $m-\lambda$ gives correct quantitative results.

We start with the simple quantum mechanical calculation.  We denote the eigenstates of the unperturbed Hamiltonian operator ($\lambda=0$) by $| l \rangle$.  With $\langle l| \hat{p}^2 | l \rangle = l+\frac{1}{2}$,
we obtain the thermal expectation value
\begin{equation} \label{eq:p2_exact}
\langle \hat{p}^2\rangle = \frac{ \sum_{l=0}^\infty\, l\; e^{-\beta m l - \frac{1}{2}\beta\lambda l (l-1)}}{ \sum_{l=0}^\infty\, e^{-\beta m l - \frac{1}{2}\beta\lambda l (l-1)}}+\frac{1}{2}.
\end{equation}
Due to the fast convergence of the two sums, this can easily be evaluated numerically.

In the functional integral, the expectation values of operators are easily computed by adding to ${\cal H}$ source terms like $K\hat{p}^2$, e.g.\ $\langle\hat{p}^2\rangle=\partial \ln Z(K)/ \partial K \big|_{K=0}$.  As long as the source terms do not involve products of $\hat{p}$ and $\hat{x}$, they simply add to the action the corresponding terms with the replacements
$f(\hat{p})\rightarrow\int d\tau\; f\left(i\left(\varphi^*-\varphi\right)/\sqrt{2}\right)$
and
$f(\hat{x})\rightarrow\int d\tau\; f\left(\left(\varphi^*+\varphi\right)/\sqrt{2}\right)$.
One obtains $\langle \hat{p}^2\rangle=-\frac{1}{2}\langle \left(\varphi^*-\varphi\right)^2\rangle$ with the usual definitions of expectation values in the functional integral
$\langle  O\!\left(\varphi,\varphi^*\right)\rangle = Z^{-1}\int {\cal D} \varphi\; O\!\left(\varphi,\varphi^*\right)\exp(-{\cal S})$.
A standard calculation in first order perturbation theory yields
\begin{equation} \label{eq:pert_res}
\langle \hat{p}^2\rangle =
\frac{1}{2} \coth \frac{\beta M}{2} - \frac{\beta\lambda\coth\frac{\beta M}{2}}{4\sinh^2\frac{\beta M}{2}},
\end{equation}
where $M=m$ corresponds to the naive coherent state approach and $M=m-\lambda$ is our suggestion.

In table~\ref{tab:results}, we show how this compares for both choices of $M$ and different values of $m$ and $\beta$ with the exact result.  We take $\lambda$ to be small against $T=1/\beta$ and $m$ in order to justify the perturbative approach.
The results clearly demonstrate that the shift in the mass term is necessary for quantitatively correct results, and this alone should be convincing that the correction we introduced to the standard functional integral is necessary for the correct interpretation of the parameters in the action.
\parbox{\linewidth}{
\begin{table}
\caption{Momentum squared $\langle \hat{p}^2\rangle$ for two different sets of $\beta$ and $m$ by exact quantum mechanics, eq.~(\ref{eq:p2_exact}), and by first order perturbation theory from the functional integral, eq.~(\ref{eq:pert_res}).  We compare the suggested shift in the mass term ($M\!=\!m\!-\!\lambda$) with the naive coherent state approach ($M\!=\!m$).
All entries must be divided by $10^3$, and the zero--temperature value of 0.5 is subtracted from $\langle \hat{p}^2\rangle$.
\label{tab:results}}
\begin{tabular}{ccccccc}
&\multicolumn{3}{c}{$\langle \hat{p}^2\rangle$ for $\beta=1,m=5$}&\multicolumn{3}{c}{$\langle \hat{p}^2\rangle$ for $\beta=3,m=1$}\\
\raisebox{1.5ex}[-1.5ex]{$\lambda$}&exact&$M\!=\!m\!-\!\lambda$&$M\!=\!m$&exact&$M\!=\!m\!-\!\lambda$&$M\!=\!m$\\
\tableline
0&6.7837&6.7837&6.7837&   52.396&52.396&52.396\\
2&6.7835&6.7835&6.7698&   52.361&52.359&52.030\\
4&6.7833&6.7832&6.7560&   52.327&52.320&51.665\\
6&6.7831&6.7830&6.7421&   52.293&52.278&51.299\\
8&6.7829&6.7827&6.7283&   52.259&52.232&50.934\\
\end{tabular}
\end{table}
}

\section{Functional integral for interacting bosons}\label{helium}

The preceeding derivation for quantum mechanics can directly be generalized to an interacting ensemble of bosons and therefore becomes relevant to the treatment of the superfluid transition of helium--4 or Bose condensation.
We describe the system through the Hamiltonian operator~(\ref{eq:H_B}), assuming the existence of a UV--momentum--cutoff $\Lambda$, $q^2\le\Lambda^2$.
The generalization of our procedure gives the action (\ref{eq:cl_ac_bosonic}) with (apart from constant terms)
$\tilde{v}(q)=v(q)$, $\tilde{\mu}=\mu+\Omega$, and
\begin{equation}\Omega=\frac{1}{2}\sum_{q^2<\Lambda^2}\bigg(\!v(0)+v(q)\!\bigg).
\end{equation}

We obtain a cutoff--dependent ``counterterm'' $\Omega$ to the chemical potential, which will be of importance for comparison with experiment.  As it turns out, this cancels a similar term generated by the one loop approximation for the ``full'' inverse propagator.  This cancellation leads to the correct low--temperature phononic dispersion relation.

We have calculated the chemical potential at the $\lambda$--transition of helium--4 under vapor pressure conditions by means of a truncation of an exact renormalization group equation \cite{Wetterich}.  The complete results are presented elsewhere \cite{GW}.  Here we just refer to the value of the critical chemical potential, which comes out to be $-6.75$~K in good agreement with the experimental value of around $-7.4$~K \cite{Maynard}.  As $\Omega$ in our case has a value of approximately 12~K (with a cutoff given by the atomic length scale), neglecting this correction would have resulted in a strong deviation from experimental values.

\section{Conclusion}

In this letter, we advocate a specific approach to the derivation of the functional integral for an interacting quantum model and its generalization to interacting bosonic systems.  We are led to a shift in the mass term of the action as compared to conventional approaches based on the use of coherent states.  A perturbative calculation of the expectation value of the squared momentum in the interacting quantum system and a renormalization group calculation of the critical chemical potential for the $\lambda$--transition of helium--4 confirm the necessity of this shift and strengthen our faith in the suggested prescription.

These findings should be important to renormalization group treatments of statistical systems such as Bose--Einstein--condensation for interacting systems as well as to the understanding of functional integrals in general.

We are grateful to Andrew Waldron for drawing our attention to the works of Christ and Lee and on the treatment of the non--linear sigma model.

\end{multicols}

\end{document}